\documentstyle[12pt]{article}
\newcommand{\eut}{
\begin{picture}(6,3)(-2,-2)
\put(1,-15){$\tilde{}$}
\put(-2,-2){$\eta$}
\end{picture}}

\newcommand{\nut}{
\begin{picture}(11,5)(0,0)
\put(5,-10){$\tilde{}$}
\put(0,0){\it N}
\end{picture}}

\newcommand{\mut}{
\begin{picture}(12,5)(0,0)
\put(6,-10){$\tilde{}$}
\put(0,0){\it M}
\end{picture}}

\begin{document}
\baselineskip=22pt plus 0.2pt minus 0.2pt
\lineskip=22pt plus 0.2pt minus 0.2pt
\begin{center}
 \Large
Homogeneous 2+1 Dimensional\\
Gravity in the Ashtekar Formulation\\

\vspace*{0.35in}

\large

J.\ Fernando\ Barbero\ G. $^{\ast,\dag}$ and
Madhavan Varadarajan $^{\ddagger}$
\vspace*{0.25in}

\normalsize

$^{\ast}$Center for Gravitational Physics and Geometry,\\
Department of Physics,\\
Pennsylvania State University,\\
University Park, PA 16802\\
U.S.A.\\

$^{\dag}$Permanent address:Instituto de Matem\'aticas\\ y F\'{\i}sica
Fundamental,
C.S.I.C.\\
Serrano 119--123, 28006 Madrid, Spain
\\

$^{\ddagger}$ Department of Physics, University of Utah, Salt lake City,
UT 84112 Utah

\vspace{.5in}
June 12, 1995\\
\vspace{.5in}
ABSTRACT
\end{center}
The constraint hypersurfaces defining the Witten and Ashtekar
formulations for 2+1 gravity are very different. In particular the
constraint hypersurface in the Ashtekar case is not a manifold but
consists of several sectors that intersect each other in a complicated
way. The issue of how to define a consistent dynamics in such a situation
is then rather non-trivial.  We discuss this point by working out the
details in a simplified (finite dimensional) homogeneous reduction of 2+1
gravity in the Ashtekar formulation.

\pagebreak

\setcounter{page}{1}

\section{Introduction}

In order to answer technical and conceptual questions which arise in the
search for a theory of quantum gravity, it is of great use to first address
these questions in the context of simpler model systems which capture some
of the features of the (more intractable) full theory. Examples
of model systems are
symmetry reductions of 3+1 gravity such as the cylindrical waves
\cite{cylwave}
and the Bianchi models (e.g. \cite{RySh}) and lower dimensional models
like 2+1 gravity and 1+1 dilatonic black holes \cite{1+1bh}.
The model system represented by 2+1 gravity has been extremely useful in
understanding some aspects related to the quantization of theories
invariant under space-time diffeomorphisms. Most of the work on 2+1 gravity
has been
done in its Witten \cite{Wit} or ADM \cite{moncrief} formulations.
In this paper we continue our investigation \cite{Madh} into aspects of the
Ashtekar formulation of 2+1 gravity \cite{Bengt}. We are motivated by the
progress in non-perturbative canonical quantization of 3+1 gravity
based on the reformulation of general relativity by Ashtekar \cite{Ash}
in terms of a new set of canonical variables.
The simplification brought
about by the use of the new
variables and, most importantly, their geometrical
meaning, have enhanced our understanding about various issues related to
quantization and have provided
the beginnings of a picture of Planck scale
gravitational physics. There are, however, several difficulties
that still have  to be
overcome, both at the technical and conceptual levels, and 2+1 gravity in the
Ashtekar formulation provides an excellent toy model for the 3+1 theory.

A turning point in our understanding of the quantum
 theory of 2+1 gravity was based on the
reformulation of the classical theory by Witten in \cite{Wit} in terms of an
$ISO(2,1)$ Chern-Simon theory. At the Hamiltonian level, the
phase space can be  coordinatized by an $SO(2,1)$ connection, and
its canonically conjugate
momentum (a densitized frame field or ``triad"), see for example \cite{joe};
this is
in close analogy with the introduction of the Ashtekar variables for
3+1 gravity. The constraints of Witten's theory are the
Gauss law constraints, which generate internal $SO(2,1)$ rotations
together with the constraints expressing  the condition that
the $SO(2,1)$ connection is flat. One
can ask the question of whether there is an Ashtekar formulation for 2+1
gravity. The
answer is affirmative. As it was shown by Bengtsson \cite{Bengt}
there are constraints
analogous to the 3+1
dimensional ones that describe 2+1 gravity. The difference with respect
to the Witten
constraints is that the condition that the connection is flat is
substituted by a
vector and a scalar constraints similar to those of 3+1 gravity.
The phase space is
the same as that of the Witten formulation and it is easy to prove
that any solution to
the Witten constraints is also a solution to the 2+1 dimensional
Ashtekar constraints.
The converse, however, is not true; the 2+1 dimensional Ashtekar
constraints are a genuine
extension of the Witten constraints and so there are solutions
to them that are not
solutions to the Witten theory.

The Ashtekar formulation of 2+1 gravity (as opposed to the Witten formulation)
shares some key features with the 3+1  formulation; it has a constraint
quadratic in the momenta (the ``triads'') and (two)
diffeomorphism constraints linear in the
momenta. These features and the fact that one cannot `Witten-ize' 3+1 gravity
to get constraints independent or at most linear in momenta
lead to important technical problems in the 3+1 case; thus we firmly believe
that a better toy model than Witten's formulation is provided by the Ashtekar
formulation of 2+1 gravity.

Another reason to consider the Ashtekar formulation of 2+1 gravity is
that one
can naturally couple local matter fields to the theory while retaining
polynomiality (in terms of the gravitational variables) of the constraints
\cite{joe}.
Local matter cannot be coupled to the Witten constraints.
The interest in studying local matter coupled to 2+1 gravity is that not
only do such systems provide  infinite dimensional non-linear
toy models but they also arise as one Killing field reductions of
vacuum 3+1 gravity
\cite{geroch}.

Note that the Ashtekar formulation (both in 2+1 and 3+1 dimensions) differs
from the ADM formulation in that it allows a natural extension to degenerate
metrics. Issues related to degenerate metrics are
important for quantization attempts \cite{isham}. In fact in
Witten's formulation of 2+1 gravity, degenerate metrics play a crucial role.
We would, therefore,
like to understand more about degenerate metrics in the Ashtekar theory and
among other things, this work investigates this
 issue in the context of the simplified model
of homogeneous 2+1 gravity.

In a previous paper \cite{Madh} we discussed some of the differences
between the Witten
and Ashtekar formulations for 2+1 gravity. Among the most interesting
results of that
analysis was discovering the fact that both theories have different
numbers of  degrees of freedom for a fixed topology of the spatial slices.
This is, in
part, a consequence of the fact that the
constraint hypersurface defined by the Witten constraints is properly
contained in the
one defined by the Ashtekar ones. One of the conclusions drawn from that
analysis was the realization of the fact that the constraint
hypersurface defined by
the Ashtekar constraints is not a manifold; actually it has
a complicated structure
and consists of several pieces glued together. An important question is,
then, how to
define dynamics in this case. One of the goals of this paper
is to give a partial
answer to this in the context of a homogeneous minisuperspace of
2+1 gravity,

We will concentrate on the study of the Ashtekar
constraints in the case
when the spatial slices in the 2+1 decomposition are tori.
We will further restrict our attention to a homogeneous
model (first introduced
by Manojlovi\'c and Mikovi\'c \cite{nenad}) obtained by imposing the
requirement that the vector fields describing the two cycles of the torus be
symmetry directions of the theory.
This is similar to the study of Bianchi models in 3+1 dimensions.

Let us briefly state what we do in this paper.
We perform an exhaustive analysis of the structure of the
constraint surface of the theory. We identify possible singularities in the
constraint surface as those points where the gradients
of the constraint functions become linearly dependent.
We find that all these possible singularities are genuine (by which we
mean that the constraint surface is not a manifold
at these points). The singularities are of two types:\\
\noindent 1) Type {\bf 1}: We can relabel sectors of the constraint
surface which contain these types of singular points
by new sets of constraint
functions. Each new set of constraints defines a smooth nonsingular manifold
which is a subset of the constraint surface (thus the new sets have a maximal
set of non-vanishing gradients). This allows the
interpretation of these singularities
as the intersection of pairs of smooth manifolds.

\noindent 2) Type {\bf 2}: These are singular regions for which we
are unable to find the simple structure which we find for Type {\bf 1}
singularities, i.e. they are not at intersections of smooth manifolds.
We show the existence of Type {\bf 2} singularities by a method
outlined in section 2. \\
 The simple structure which we have been able to find for Type {\bf 1} points
enables us to study the gauge orbits in Type {\bf 1} regions and examine
issues of dynamics.  We refrain from saying anything about dynamics
in the Type
{\bf 2} case because of its more complicated character.
The main result of this work is that, if we cut out Type {\bf 2}
regions from the
phase space, the physically relevant part of the reduced phase space
of the Witten and the Ashtekar formulations of homogeneous 2+1 gravity on
the torus are identical.

The lay out of the paper
is the following. In section 2 we identify the possible singularities
of the constraint hypersurface defined by the
homogeneous Ashtekar constraints on a 3 manifold with topology $T^2 \times R$.
In section 3 we identify type {\bf 1} and type {\bf 2} singularities
and introduce the
new constraint functions that allow us to define
type {\bf 1} singularities as intersections of smooth manifolds.
 In section 4 we study the dynamics of
the model. In particular,
we  define ``physical" initial data (for which
the 2-metric is non-degenerate and has (++) signature) and describe
their evolution.  We carefully analyze the issue of how to
evolve through the Type {\bf 1} singularities of the constraint hypersurface.
We end the paper with
our conclusions and some speculations in section 5.

\section{The Constraint Hypersurface}

This section is devoted to the description of the constraint hypersurface
for homogeneous 2+1 gravity in the Ashtekar formulation and the study of its
singularities. In order to
describe a constrained Hamiltonian system, the first step is the
introduction of the phase space $\Gamma$, an
even dimensional manifold\footnote{We
will denote the coordinates in (a chart of) $\Gamma$ as
$\{x^{\alpha}\}$} with a symplectic structure given by a 2-form  $\Omega\equiv
\Omega_{\alpha\beta}dx^{\alpha}\wedge dx^{\beta}$ defined on it. There are two
conditions that $\Omega$ must satisfy. First, it must be closed, that is,
$d\Omega=0$. This closure condition is necessary in order to guarantee
that the Poisson brackets will satisfy the Jacobi identity. Second, it
must be non-degenerate, that is,
$\Omega_{\alpha\beta}v^{\beta}=0\Leftrightarrow
v^{\alpha}=0$. The non-degeneracy of $\Omega$ means that
it is possible (though some
subtleties apply for phase spaces of infinite dimension) to
define the inverse $\Omega^{\alpha\beta}$ as
$\Omega^{\alpha\beta}\Omega_{\beta\gamma}=-\delta^{\alpha}_{\gamma}$. With
its aid we can define the Poisson bracket of any pair of
functions $f$ and $g$ in
$\Gamma$ as
\begin{equation}
\{f,g\}\equiv \Omega^{\alpha\beta}\partial_{\alpha}f
\partial_{\beta}g
\label{xyz}
\end{equation}
where
$\partial_{\alpha}$ is a torsion-free derivative operator.

In order to describe a constrained Hamiltonian system we need to add
constraints. These are conditions that the dynamical variables must
satisfy; they are given by functions in the phase space
$C_{i}(x)=0; \;\;i=1,\ldots, P$. A set of constraints is said to be first
class if the Poisson brackets of any two of them is zero on the constraint
hypersurface. This is equivalent to the condition $\{C_{i},
C_{j}\}=f^{k}_{ij}C_{k}$ where the $f^{k}_{ij}$ are antisymmetric in
$i,\;j$ and, possibly, coordinate dependent. The definition of first class
constraints admits the following interpretation. The functions $C_{i}$
define a hypersurface $\gamma$ immersed in $\Gamma$. The definition of
first class constraints introduced above means that if we take any
normal to $\gamma$ (given by a linear combination of the gradients of the
constraint functions $dC_{i}=\partial_{\alpha}C_{i}dx^{\alpha}$) and build
the vector
field $S^{\alpha}_{i}\equiv\Omega^{\alpha\beta}\partial_{\beta}C_{i}$ then
$S^{\alpha}_{i}$ is tangent to $\gamma$. These vector fields tangent to the
constraint hypersurface can be integrated to get the gauge orbits on
$\gamma$ whose points describe physically equivalent configurations of the
system. In the rest of the paper we will use this geometrical
interpretation for first class constraints. One of the issues that we want
to emphasize from the beginning is that once the constraint hypersurface is
given, the specific functions $C_{i}$ introduced in order to define it are
irrelevant.  All the steps in the definition of a first class system can
be justified in purely geometrical terms without having to consider any
explicit form of the constraint functions. In some cases when pathologies
in the definitions of gauge orbits etc. appear, they can be traced back
to the vanishing of the gradients of some of the $C_{i}$ or to the fact that
some of these gradients become linearly dependent.
In these situations a genuine pathology
may be present; the hypersurface $\gamma$ may have some sort of
singularity that makes it impossible to define gauge orbits in a
consistent way. It may happen, though, that the problem is caused by a bad
choice of the constraint functions and not by the hypersurface itself,
which may be smooth and perfectly well behaved.
If this is the case, a judicious
choice of $C_{i}$ in the vicinity of the points of $\gamma$ where the
problem appears may be enough to circumvent it. Even if genuine
singularities are present, it may still be possible to define a consistent
dynamics by considering, for example, gauge orbits that are not manifolds
but such that the reduced phase space is. The geometrical point of view
that we will adopt in this paper can be summarized by saying
that ``only the constraint
hypersurface matters". The vanishing of the
gradients of the constraint functions
must be taken as a warning sign but the presence or absence of
singularities has to be
carefully studied.

A  trivial but illustrative example of the above is the following.
Consider the
circumference $S^{1}$ as defined on ${\rm l\!R^{2}}$ by
$F(x,y)=x^{2}+y^{2}-1=0$. The gradient $dF=2(xdx+ydy)$ is non-zero for all
the points in $S^{1}$ and then this is a smooth manifold. We could have
used the function $G(x,y)=(x^{2}+y^{2}-1)^{2}=0$ to describe the same
circumference, instead, but now $dG=2(x^{2}+y^{2}-1)(xdx+ydy)$ is zero for
all the points of $S^{1}$. The vanishing of $dG$ does not signal any
problem with $S^{1}$ but, rather, that the choice of functions to describe
it is not very clever. An example in which the vanishing of a gradient
really implies the existence of a singularity is the cone
$F(x,y,z)=z^2-x^2-y^2=0$. At the vertex $(0,0,0)$ the gradient
$dF=2(zdz-xdx-ydy)$ is zero. In order to show that the point $(0,0,0)$ is
indeed a
singularity we check that the tangent space there is not well defined.
To this end we
take three curves contained in the cone parametrized as $\gamma_{1}\equiv
(\lambda, 0 ,\lambda)$, $\gamma_{2}\equiv (-\lambda, 0 ,\lambda)$,
$\gamma_{3}\equiv
(0,\lambda, \lambda, )$ and compute the tangent vectors at $(0,0,0)$
(rather take the
limit of the tangent vectors as the points approach $(0,0,0)$ ).We find that
the tangent vectors
$\tau_{1}=(1,0,1)$, $\tau_{2}=(-1,0,1)$, $\tau_{3}=(0,1,1)$  are linearly
independent. In any other point P of the cone (where it is a locally a
two-dimensional manifold) if
three curves intersect then the three corresponding tangent
vectors are linearly
dependent. Thus the presence of extra linearly independent vectors at
$(0,0,0)$
signals that this point is a genuine singularity.

A model in which all the issues discussed above are relevant is 2+1
gravity in the Ashtekar formulation. This is an interesting system
because it is
possible to find
different sets of first class constraints that describe several
(at times overlapping) regions of the constraint hypersurface \cite{Madh}.
The issue of the
compatibility of the dynamics
defined by the different sets of constraints arises, as
well as the appearance of singularities. In the rest of the paper we
will discuss a
simplified version of 2+1 gravity in the Ashtekar formulation.
We will concentrate on a homogeneous case where the
2-slices are tori, in which the
fields can be taken as coordinate independent. In spite of the
simplification that this
entails the system keeps several interesting features that make it worth
studying;
(remember, for example, that on the torus, the Witten constraints
define an essentially
homogeneous model).

We give now our conventions and notation. The configuration
variable for 2+1 gravity is a real $SO(2,1)$ valued connection $A_{a}^{I}$
with conjugate momentum $\tilde{E}^{a}_{I}$ (the frame fields or ``triads").
In the following a, b, c, etc.
(running from 1 to 2) will represent tangent space indices; internal indices
will be denoted by I, J, K, etc (running from 1 to 3). They are raised and
lowered with the (internal) Minkowski metric $\eta_{IJ}$ with signature
(--, +,
+). The Levi-Civita
tensor density and its inverse will be denoted as $\tilde{\eta}^{ab}$ and
$\eut_{ab}$ respectively. The convention of representing the
density weight of an object with tildes above or below the fields (positive
and negative density weights respectively) will be used throughout the paper.
The covariant derivatives are given by
$\nabla_{a}\alpha_{I}=\partial_{a}\alpha_{I}+\epsilon_{IJ}^{\;\;\;\;K}
A_{a}^{J}\alpha_{K}$, the curvature is $F_{abI}=2\partial_{[a}A_{b]I}+
\epsilon_{I}^{\;\;JK}A_{aJ}A_{bK}$, where $\epsilon^{IJK}$ is the internal
Levi-Civita tensor ($\epsilon^{123}=1$) and finally the Poisson brackets
between the
connection and frame fields are $\{A_{a}^{I}(x), \tilde{E}^{b}_{J}(y)\}=
\delta^{2}(x,y)\delta^{\;\;b}_{a}\delta^{\;\;I}_{J}$

\noindent  The Witten constraints for 2+1 gravity are \cite{Wit}:
\begin{eqnarray}
& & \nabla_{a} \tilde{E}^{a}_{I}=0\nonumber\\
& & F_{ab}^{I}=0\label{j1}
\end{eqnarray}
whereas the Ashtekar constraints in this case  are \cite{Bengt}:
\begin{eqnarray}
& & \nabla_{a} \tilde{E}^{a}_{I}=0\nonumber\\
& & \tilde{E}^{b}_{I} F_{ab}^{I}=0\label{j2}\\
& & \epsilon^{IJK}\tilde{E}^{a}_{I}\tilde{E}^{b}_{J}F_{abK}=0\nonumber
\end{eqnarray}

\noindent They are called the Gauss, vector and scalar (or Hamiltonian)
constraints
respectively. Both (\ref{j1}) and (\ref{j2}) are first class systems
and are equivalent when the triads are non--degenerate \cite{Bengt}.
In contrast with the more familiar 3+1 dimensional case the variables used in
(\ref{j2}) are real and thus no reality conditions need to be included in the
formalism.
The fact that we have six constraints and six configuration variables
per point
indicates, via naive counting, that there may be topological but no
local degrees of
freedom.

In homogeneous models it
is always possible to introduce
bases of vectors and one-forms in such a way that the
partial derivatives of the fields can be traded for
expressions involving the structure
constants of the isometry group. In our case, this will be chosen to be
the 2-dimensional
 group  $U(1)\times U(1)$ whose
abelian character implies the vanishing of the structure
constants. This means that we can remove the derivatives in the
definitions introduced above, and the Poisson brackets
between the dynamical variables become $\{A_{a}^{I}, \tilde{E}^{b}_{J}\}=
\delta^{\;\;b}_{a}\delta^{\;\;I}_{J}$

All the systems of constraints that we will use in this paper
share in common the Gauss law
that for homogeneous fields is
\begin{equation}
{\tilde G}_{I}\equiv\epsilon_{I}^{\;\;JK}A_{aJ}{\tilde E}^{a}_{K}=0
\label{002}
\end{equation}
We will discuss it carefully before introducing any other constraints.
In the following arguments it is
very convenient to think of the fields $A_{a}^{I}$ and ${\tilde E}^{a}_{I}$
as the
components of four
$SO(2,1)$ vectors $A_{1\,I}$, $A_{2\,I}$, ${\tilde E}^{1}_{I}$, and ${\tilde
E}^{2}_{I}$ because it will be usually possible to understand the
meaning of algebraic
statements on them as some simple geometrical relationship between
these 3 dimensional
objects. The Gauss law, for example, can be interpreted as the condition
\begin{equation}
A_{1}\times
{\tilde E}^{1}+A_{2}\times {\tilde E}^{2}=0
\label{002bis}
\end{equation}
where the vector product $(A \times B)_{I}$ is defined by $(A \times
B)_{I}\equiv\epsilon_{I}^{\;\;JK} A_{J} B_{K}$ (notice
that the first index in $\epsilon_{I}^{\;\;JK}$ is lowered with the
Minkowski metric
$\eta_{IJ}$). It has
properties analogous to those of the vector product in
${\rm l\!R}^3$; for example the
vector product of two $SO(2,1)$ vectors is normal, in the Lorentz sense,
to the two vectors themselves. A
consequence of this is that the Gauss law requires that
$A_{1\,I}$, $A_{2\,I}$, ${\tilde E}^{1}_{I}$ and ${\tilde
E}^{2}_{I}$ must be linearly dependent, i.e. contained in the same plane.
This is so
because (\ref{002bis}) tells us that  $A_{1}\times {\tilde E}^{1}$ must be
proportional to $A_{2}\times {\tilde E}^{2}$ and then
the planes containing both
couples of vectors must coincide. Generically we can freely specify three of
these
vectors in  this plane and have a one parameter freedom to choose the fourth.

We now look at the
gradients $d{\tilde G}^{I}$; we will need them in order to study the possible
singularities of the constraint manifold defined by the homogeneous Ashtekar
constraints that we will introduce later. We have
\begin{equation}
d{\tilde G}^{I}=\epsilon^{IJK}A_{aJ}d{\tilde
E}^{a}_{K}-\epsilon^{IJK}{\tilde E}^{a}_{J}dA_{aK}\equiv J\left[
\begin{array}{l}
d{\tilde E}^{a}_{I} \\
dA_{aI}
\end{array}
\right]
\label{002ter}
\end{equation}
Where J is a $3\times 12$ matrix
\begin{equation}
\left[
\begin{array}{rrrrrrrrrrrr}
\!\!0 & \!\!-A_{1\,3} & \!\!A_{1\,2} & \!\!0 & \!\!-A_{2\,3} &
\!\!A_{2\,2} & \!\!0 &
\!\!{\tilde E}
^{1}_{3} &
\!\!-{\tilde E}^{1}_{2} &  \!\!0 & \!\!{\tilde E}^{2}_{3} &
\!\!-{\tilde E}^{2}_{2} \\
\!\!A_{1\,3} & \!\!0 & \!\!-A_{1\,1} &\!\! A_{2\,3} &\! \!0 &
\!\!-A_{2\,1} & \!\!-
{\tilde E}^{1}_{3} &\! \!0 &
\!\!{\tilde E}^{1}_{1} & \!\!-{\tilde E}^{2}_{3} & \!\!0 &
\!\!{\tilde E}^{2}_{1} \\
\!\!-A_{1\,2} & \!\!A_{1\,1} & \!\!0 & \!\!-A_{2\,2} & \!\!A_{2\,1} &
\!\!0 & \!\!
{\tilde E}^{1}_{2} &
 \!\!-{\tilde E}^{1}_{1} & \!\!0 & \!\!{\tilde E}^{2}_{2} &
\!\!-{\tilde E}^{2}_{1}
& \!\!0
\end{array}
\right]
\label{002tetr}
\end{equation}
(the second index in the components of the connection is the internal index)
The gradients of the three functions ${\tilde G}^{I}$ will be linearly
independent if
and only
if the rank of (\ref{002tetr}) is 3 for connections and ``triads"
satisfying the
Gauss law. A necessary and sufficient condition for this to happen is
that any two of the
four internal vectors $A_{1I}$, $A_{2I}$, ${\tilde E}^{1}_{I}$ and ${\tilde
E}^{2}_{I}$ (satisfying the Gauss law) are linearly independent as we show in
the following paragraphs.

 From the form
of (\ref{002tetr}) it is clear
that if  we have a non zero vector among the
 $(A_{a}^{I},{\tilde E}^{b}_{J})$, then the rank of the matrix
is, at least, two. Without loss of generality we can choose this vector to be
$A_{1\,I}$. We consider now a linear change of coordinates
in $\Gamma$ given by
\begin{equation}
\left[
\begin{array}{c}
{\tilde E}^{1\;*}_{I}\\
{\tilde E}^{2\;*}_{I}\\
A^{*}_{1\,I}\\
A^{*}_{2\,I}
\end{array}
\right]=\left[
\begin{array}{cccc}
S_{I}^{\;\;J} & 0 & 0 & 0 \\
0 & S_{I}^{\;\;J} & 0 & 0 \\
0 & 0 & S_{I}^{\;\;J} & 0 \\
0 & 0 & 0 & S_{I}^{\;\;J}
\end{array}
\right]
\left[
\begin{array}{c}
{\tilde E}^{1}_{J}\\
{\tilde E}^{2}_{J}\\
A_{1\,J}\\
A_{2\,J}
\end{array}
\right]
\label{003}
\end{equation}
where $ S_{I}^{\;\;J}$ is a constant, non-singular matrix
(i.e. independent of $A_{a}^{I}$ and
${\tilde E}^{a}_{I}$ ) belonging to $Gl(3,{\rm I}\!{\rm R})$.
Under this transformation the
gradient of the Gauss law becomes
\begin{equation}
d{\tilde G}^{I}=\frac{1}{\det S}S_{J}^{\;\;I}d{\tilde G}^{J\,*}
\label{004}
\end{equation}
The gradient matrix (\ref{002tetr})
consists of four $3\times 3$ antisymmetric square
boxes that transform with the same matrix $S_{I}^{\;\;J}$ under (\ref{003}).
It is
always possible to find a matrix $ S_{I}^{\;\;J}$ (belonging to $SO(3)$)
in such a way that one of these boxes takes the form
\begin{equation}
\left[
\begin{array}{ccc}
0 & \alpha & 0 \\
-\alpha & 0 & 0\\
0 & 0 & 0
\end{array}
\right]
\end{equation}
so that, without loss of generality, we can write $A_{1\,3}\neq 0$ ,
$A_{1\,2}=0$ and
$A_{1\, 1}=0$. If there is another vector that is not
collinear with $A_{1}^{I}$, such
that the Gauss law is satisfied, then at least one of $A_{2\,2}$,
$A_{2\,1}$,
${\tilde E}^{1}_{2}$, $\!-{\tilde E}^{1}_{1}$, $\!{\tilde E}^{2}_{2}$,
$\!-{\tilde E}^{2}_{1}$ must be different from zero and then the rank of the
matrix is
obviously 3. If the four internal vectors are collinear (which is
trivially a
solution to the Gauss law), proportional to $A_{1\,I}$, and we  write it as
before
with  $A_{1\,3}\neq 0$, $A_{1\,2}=0$ and $A_{1\,1}=0$ we see that now
$A_{2\,2}$,
$A_{2\,1}$, ${\tilde E}^{1}_{2}$, $\!-{\tilde E}^{1}_{1}$,
$\!{\tilde E}^{2}_{2}$,
$\!-{\tilde E}^{2}_{1}$ are all zero, and then the rank of (\ref{002tetr})
is only 2.
We conclude that we expect to find singularities in the hypersurface
defined by the Gauss law when the four vectors $A_{aI}$ and ${\tilde
E}^{a}_{I}$ are all collinear.

The main purpose of this paper is to study the system of constraints given by
${\tilde G}_{I}=0$ and the homogeneous version of the Ashtekar Hamiltonian
constraint $\;\;{\tilde{\!\!\tilde A}}=0$; where
\begin{equation}
{\tilde{\!\!\tilde A}}\equiv{\tilde v}_{I}{\tilde w}^{I}
=-2[({\tilde E}^{a}_{I}A_{a}^{I})
    ({\tilde E}^{b}_{J}A_{b}^{J})-({\tilde E}^{a}_{I}A_{b}^{I})
    ({\tilde E}^{b}_{J}A_{a}^{J})]\label{008}
\end{equation}
and
\begin{equation}
\begin{array}{l}
{\tilde v}_{I}=\epsilon_{I}^{\;\;JK}{\tilde \eta}^{ab}A_{aJ}A_{bK}\\
{\tilde w}_{I}=\epsilon_{I}^{\;\;JK}\eut_{ab}{\tilde E}^{a}_{J}
{\tilde E}^{b}_{K}
\end{array}
\label{008bis}
\end{equation}
In order to get (\ref{008}) we have used the fact that, in the homogeneous
case that
we are considering in this paper, the curvature $F_{ab}^{I}$ is given by
\begin{equation}
F_{ab}^{I}\equiv\epsilon_{I}^{\;\;JK}A_{aJ}A_{bK}
\label{008ter}
\end{equation}
Notice that the vector constraint disappears in this case because it is
always proportional to the Gauss law (${\tilde E}^{a}_{I}F_{ab}^{I}=
\epsilon^{IJK}
{\tilde E}^{a}_{I}A_{aJ}A_{bK}={\tilde G}^{K}A_{bK}=0$).

We need to study now the rank
of the $4\times 12$ matrix K defined by
\begin{equation}
\left[
\begin{array}{c}
d\;{\tilde{\!\!\tilde A}}\\
d{\tilde G}^{I}
\end{array}
\right]\equiv
\left[
\begin{array}{c}
K\\J
\end{array}
\right]
\left[
\begin{array}{c}
d{\tilde E}^{a}_{I}\\
dA_{aI}
\end{array}
\right]
\label{102}
\end{equation}
where
\begin{equation}
d\;{\tilde{\!\!\tilde A}}=2\epsilon^{IJK}\left( \eut_{ab}{\tilde v}_{I}
{\tilde E}^{a}_{J}d{\tilde E}^{b}_{K}+{\tilde \eta}^{ab}{\tilde
w}_{I}A_{aJ}dA_{bK}\right)\equiv K \left[
\begin{array}{c}
d{\tilde E}^{a}_{I}\\
dA_{aI}
\end{array}
\right]
\label{103}
\end{equation}
It is straightforward
to show that, whenever the rank of the matrix J (defining $d{\tilde G}_{I}$)
is not maximal, both ${\;\;\tilde{\!\!\tilde A}}=0$ and ${\tilde G}_{I}=0$.
We conclude,
then, that all points in the constraint hypersurface such that
the four internal vectors
$A_{a I}$ and ${\tilde E}^{a}_{I}$ are collinear are possible singularities.
We will
restrict ourselves now to configurations such that
$d{\tilde G}_{I}$ has maximal rank. There are three different cases to
consider
according to the time-like, space-like or null character of the normal to
the plane
containing $A_{a I}$ and ${\tilde E}^{a}_{I}$. The result of a detailed
analysis that
follows the same lines as the discussion of the Gauss law made above
shows that in all
these three cases we have possible singularities whenever $A_{aI}=0$ or
${\tilde
E}^{a}_{I}=0$ or both $A_{aI}$ are linearly dependent or both
${\tilde E}^{a}_{I}$ are
linearly dependent. In the case when the plane that contains $A_{a I}$ and
${\tilde E}^{a}_{I}$ is null it is not necessary to have  $A_{1I}$
and $A_{2I}$
linearly dependent in order to solve the Gauss law; we have then additional
possibly singular configurations that we describe in some detail now
(a similar
situation occurs if we interchange
the roles of $A_{a I}$ and
${\tilde E}^{a}_{I}$ ). By using an $SO(2,1)$
transformation  we can always write ($\alpha\neq 0$)
\begin{equation}
\begin{array}{c}
A_{1I}=(0, \alpha, 0)\\
A_{2I}=(1, \beta, 1)\\
{\tilde E}^{1}_{I}=(\gamma, \delta, \gamma)\\
{\tilde E}^{2}_{I}=(\epsilon, \theta, \epsilon)
\end{array}
\label{115}
\end{equation}
and then $\epsilon_{I}^{\;\;JK}A_{1J}{\tilde
E}^{1}_{K}=(-\alpha\gamma,0,-\alpha\gamma)$ and
$\epsilon_{I}^{\;\;JK}A_{2J}{\tilde
E}^{2}_{K}=(-\beta\epsilon+\theta,0,-\beta\epsilon+\theta)$.
Notice that the fact that $d{\tilde G}_{I}$ has maximal rank implies that
at least one
vector (that we choose to be $A_{1}$) is not null.
The Gauss law tells us
that $\alpha\gamma+\beta\epsilon=\theta$. The matrix $K$ is
\begin{equation}
\left[
\begin{array}{cccccccccccc}
-4\alpha\theta&0&4\alpha\theta&4\alpha\delta&0&-4\alpha\delta&4\beta\sigma&0&
-4\beta\sigma&-4\alpha\sigma&0&4\alpha\sigma\\
0&0&\alpha&0&-1&\beta&0&\gamma&-\delta&0&\epsilon&-\theta\\
0&0&0&1&0&-1&-\gamma&0&\gamma&-\epsilon&0&\epsilon\\
-\alpha&0&0&-\beta&1&0&\delta&-\gamma&0&\theta&-\epsilon&0
\end{array}
\right]
\label{116}
\end{equation}
The rank of this matrix will be maximal if and only if
$\sigma\equiv\epsilon\delta-\gamma\theta\neq 0$;
i.e. if and only if ${\tilde E}^{1}_{I}$ and ${\tilde E}^{2}_{I}$ are
not collinear. Notice that $\theta$ and $\delta$ may be both different
from zero, in
which case we have that the rank of $K$ is 3 with the gradient of the scalar
constraint different from zero. This is in contrast with the types of
singularities encountered before, for which $d\;{\tilde{\!\!\tilde A}}=0$.

We summarize the possible singularities of the constraint hypersurface (see
figure 1) defined by
the homogeneous Ashtekar constraints. In all the cases considered above
we have
singularities if

\noindent {\bf a} $A_{1\,I}$, $A_{2\,I}$, ${\tilde E}^{1}_{I}$,
and ${\tilde E}^{2}_{I}$ are
all collinear. These are the singularities of the Gauss law.

\noindent {\bf b} ${\tilde E}^{a}_{I}=0$ and the $A_{a\,I}$ linearly
independent but, otherwise,
arbitrary.

\noindent {\bf c} $A_{a\,I}=0$ and ${\tilde E}^{a}_{I}$  linearly independent
but, otherwise, arbitrary.

\noindent {\bf d} $(A_{1\,I}, A_{2\,I})$ are linearly dependent
and $({\tilde E}^{1}_{I}, {\tilde
E}^{2}_{I})$ are also linearly dependent but not collinear
with $(A_{1\,I}, A_{2\,I})$.
In this case both ${\tilde v}^{I}$ and
${\tilde w}^{I}$ are zero.

In all the previous cases we have that $d\;{\tilde{\!\!\tilde
A}}=0$. In addition to these, if the plane containing $A_{aI}$
and ${\tilde E}^{a}_{I}$
is null, then there are also possible singularities in two other situations:

\noindent {\bf e} ${\tilde E}^{1}_{a}$ and ${\tilde E}^{2}_{a}$ are
linearly dependent
with $A_{1I}$, $A_{2I}$ contained in the null plane, non-collinear
but, otherwise, arbitrary

\noindent {\bf f} $A_{1\,I}$ and $A_{2\,I}$ are linearly dependent;
${\tilde E}^{1}_{I}$ and ${\tilde E}^{2}_{I}$ are contained in the
null plane (but
are otherwise arbitrary) and
non-collinear.

Notice that some of these last configurations are such that we have
possible singularities
in spite of having $d\;{\tilde{\!\!\tilde A}}\neq 0$.
In the previous classification
we have excluded from a certain type those configurations
that can be classified in a
previous type. For example, we have excluded from {\bf e} those
configurations with  $A_{1\,I}$ and $A_{2\,I}$ collinear and
classified them as type
{\bf d}
\begin{figure}
\begin{picture}(500,250)(0,0)

\put(7,180){\vector(3,1){25}}
\put(7,180){\vector(3,1){50}}
\put(7,180){\vector(3,1){75}}
\put(7,180){\vector(3,1){100}}
\put(60,170){{\bf a}}
\put(30,194){$A_{1}$}
\put(55,203){$A_{2}$}
\put(79,209){${\tilde E}^{1}$}
\put(104,217){${\tilde E}^{2}$}
\put(7,180){\circle*{3}}

\put(196,200){\vector(3,-1){60}}
\put(196,200){\vector(4,3){40}}
\put(221,170){{\bf b}}
\put(216,230){$A_{1}$}
\put(220,193){$A_{2}$}
\put(155,197){${\tilde E}^{a}=0$}
\put(196,200){\circle*{3}}

\put(361,200){\vector(3,-1){60}}
\put(361,200){\vector(4,3){40}}
\put(386,170){{\bf c}}
\put(381,226){${\tilde E}^{1}$}
\put(385,193){${\tilde E}^{2}$}
\put(320,197){$A_{a}=0$}
\put(361,200){\circle*{3}}

\put(10,80){\vector(3,1){45}}
\put(10,80){\vector(3,1){100}}
\put(10,80){\vector(1,-4){8}}
\put(10,80){\vector(-1,4){9}}
\put(60,45){{\bf d}}
\put(21,44){$A_{1}$}
\put(05,113){$A_{2}$}
\put(45,100){${\tilde E}^{1}$}
\put(104,117){${\tilde E}^{2}$}
\put(10,80){\circle*{3}}

\put(150,80){\vector(4,1){45}}
\put(150,80){\vector(4,1){100}}
\put(150,80){\vector(2,-1){30}}
\put(150,80){\vector(1,3){9}}
\put(140,77){\line(3,1){120}}
\put(140,77){\line(-1,2){15}}
\put(140,77){\line(1,-2){15}}
\put(260,117){\line(-1,2){15}}
\put(260,117){\line(1,-2){15}}
\put(125,107){\line(3,1){120}}
\put(155,47){\line(3,1){120}}
\put(215,45){{\bf e}}
\put(161,59){$A_{1}$}
\put(161,105){$A_{2}$}
\put(185,75){${\tilde E}^{1}$}
\put(244,89){${\tilde E}^{2}$}
\put(150,80){\circle*{3}}

\put(305,80){\vector(4,1){45}}
\put(305,80){\vector(4,1){100}}
\put(305,80){\vector(2,-1){30}}
\put(305,80){\vector(1,3){9}}
\put(295,77){\line(3,1){120}}
\put(295,77){\line(-1,2){15}}
\put(295,77){\line(1,-2){15}}
\put(415,117){\line(-1,2){15}}
\put(415,117){\line(1,-2){15}}
\put(280,107){\line(3,1){120}}
\put(310,47){\line(3,1){120}}
\put(370,45){{\bf f}}
\put(316,59){${\tilde E}^{1}$}
\put(316,105){${\tilde E}^{2}$}
\put(340,75){$A_{1}$}
\put(399,89){$A_{2}$}
\put(305,80){\circle*{3}}

\end{picture}
\caption{Possible singularities of the constraint hypersurface.
In {\bf a}, {\bf b},
{\bf c} and {\bf d}
the plane containing $A_{aI}$ and ${\tilde E}^{a}_{I}$ is arbitrary,
whereas in {\bf e}
and {\bf f} it must be null.}
\end{figure}
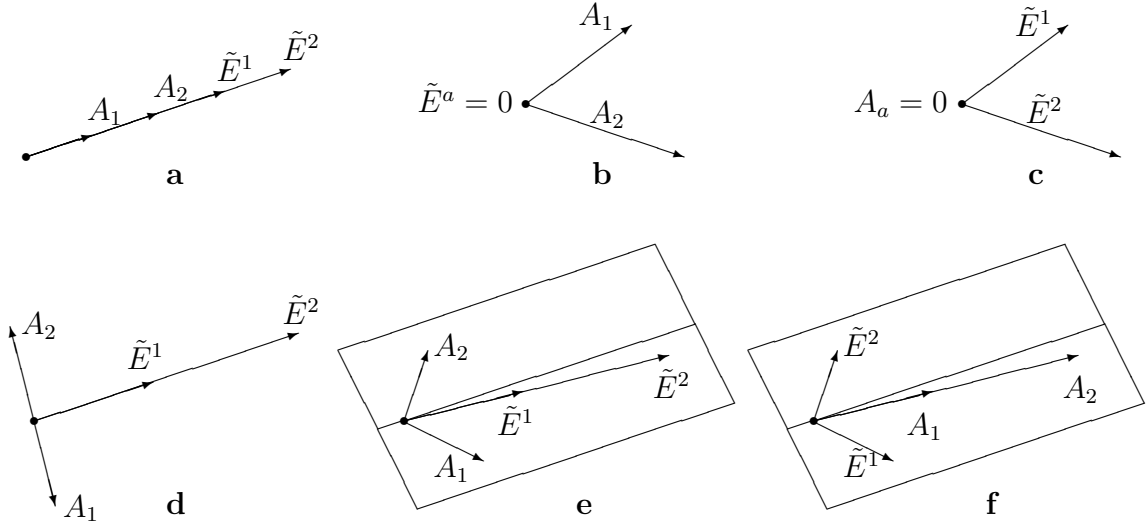

Although it is possible to check at this point that all the previously
described field
configurations are indeed singularities of the constraint hypersurface
by explicitly
showing that the tangent space is not defined there (as we did with the
example of
the cone) we will follow a different strategy. As we shall see
in the following, it is
possible to describe some parts of the constraint hypersurface
with constraint
functions different from $\;{\tilde{\!\!\tilde A}}=0$
and ${\tilde G}^{I}=0$. The
possible singularities of the hypersurfaces (sectors of the full constraint
hypersurface) defined by these new sets of constraints can be identified
proceeding as
before. It turns out that some of the configurations shown in fig. 1
are not singular
for some of these new sets of constraints. However, in these cases,
it turns out that the relevant part of
the constraint hypersurface is an intersection of two smooth hypersurfaces
(defined by the new sets of constraint functions)
in the phase space that are strictly
contained in  $\;\;{\tilde{\!\!\tilde A}}=0$ and ${\tilde G}^{I}=0$.
These ``intersection'' type of
singularities will be referred to as type {\bf 1}.
We will see that it is possible to
define dynamics in a consistent way even if they are present. The remaining
singularities will be called type {\bf 2}.

\section{Singularities and New Constraints}

The starting point of this section is the observation of the fact that in the
non-homogeneous case \cite{Madh} there are systems of first class
constraints that
extend the Witten ones but describe only some of the sectors
present in the Ashtekar
formulation. When we specialize these new constraints to
the homogeneous case we are
led to consider several systems of first class constraints consisting
of the Gauss law
and any of the following functions
\begin{eqnarray}
& & {\tilde v}_{I}\equiv
    \tilde{\eta}^{ab}\epsilon_{IJK}A_{a}^{J}A_{b}^{K}=0\label{004bis}\\
& & {\tilde w}_{I}\equiv \eut_{ab}\epsilon_{IJK}{\tilde E}^{aJ}
{\tilde E}^{bK}=0
    \label{005}\\
& & {\;\;\tilde {\!\!\tilde M}}\equiv{\tilde v}_{I}{\tilde v}^{I}
=-2{\tilde \eta}^{ab}
    {\tilde \eta}^{cd}(A_{a}^{I}A_{cI})(A_{b}^{J}A_{dJ})=0\label{006}\\
& & {\;\;\tilde{\!\!\tilde F}}\equiv {\tilde w}_{I}{\tilde w}^{I}
=-2\eut_{ab}\eut_{cd}
    ({\tilde E}^{a}_{I}{\tilde E}^{cI})({\tilde E}^{b}_{J}{\tilde E}^{dJ})
=0\label{007}
\end{eqnarray}
Whereas in the non-homogeneous case the roles of connections and triads
are very
different, in the present situation we find a curious duality:
the homogeneous
version of the Ashtekar constraints  is invariant under
the interchange of $A_{aI}$ and
${\tilde E}^{a}_{I}$. As a consequence of this,
any statement made for a particular
set of phase space points    will have an analog in which the role of the
connection and ``triad''   is
interchanged.
\noindent The gradients of (\ref{004bis}-\ref{007}) are given by
\begin{eqnarray}
& & d{\tilde v}^{I}=2{\tilde \eta}^{ab}\epsilon_{IJK}A_{a}^{J}
dA_{b}^{K}\label{010}\\
& & d{\tilde w}^{I}=2\eut_{ab}\epsilon^{IJK}{\tilde E}^{a}_{J}d{\tilde
E}^{b}_{K}\label{011}\\
& & d{\;\tilde {\!\!\tilde M}}=4{\tilde
v}^{I}{\tilde\eta}^{ab}\epsilon_{IJK}A_{a}^{J}dA_{b}^{K}\label{013}\\
& & d{\;\tilde {\!\!\tilde F}}=4{\tilde
w}^{I}\eut_{ab}\epsilon_{IJK}{\tilde E}^{a}_{J}d{\tilde E}^{b}_{K}\label{014}
\end{eqnarray}

In all these cases there are only four independent constraint
equations regardless of the fact that some of the additional constraints
are internal
vector densities; in other words, all the systems of constraints that
we will consider define
hypersurfaces of the same dimensionality in the phase space.
Recall  that the analog of the vector
constraints in the non-homogeneous case do not appear here because they are
always
proportional to the Gauss law.

The relationship between
all these different systems of constraints (or rather how the different
constraint
hypersurfaces defined by them are contained in each other)
is summarized in figure 2.
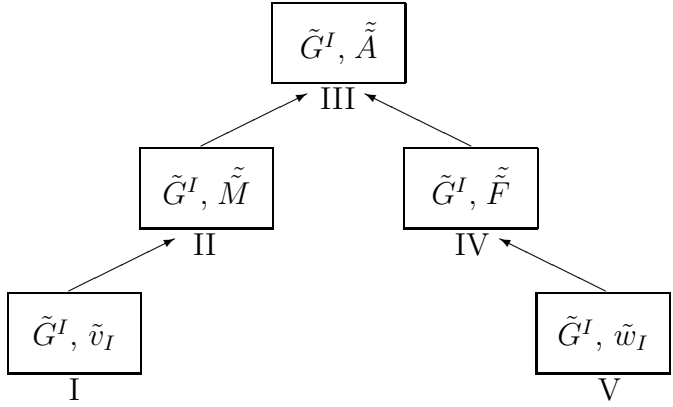
\begin{figure}
\begin{picture}(500,150)(0,0)
\put(80,10){\framebox(50,30){${\tilde G}^{I}$, ${\tilde v}_{I}$}}
\put(103,0){I}
\put(280,10){\framebox(50,30){${\tilde G}^{I}$, ${\tilde w}_{I}$}}
\put(303,0){V}
\put(130,65){\framebox(50,30){${\tilde G}^{I}$, $\;{\tilde{\!\!\tilde M}}$}}
\put(150,55){II}
\put(230,65){\framebox(50,30){${\tilde G}^{I}$, $\;{\tilde{\!\!\tilde F}}$}}
\put(249,55){IV}
\put(180,120){\framebox(50,30){${\tilde G}^{I}$, $\;{\tilde{\!\!\tilde A}}$}}
\put(198,110){III}
\put(103,41){\vector(2,1){40}}
\put(153,96){\vector(2,1){40}}
\put(306,41){\vector(-2,1){40}}
\put(255,96){\vector(-2,1){40}}
\end{picture}
\caption{Relationship between the different systems of constraints.}
\end{figure}

The fact that $I\subset II$, $I\subset III$, $V\subset IV$, and $V\subset III$
 is trivial. Notice, however, that ${\tilde v}^{I}{\tilde v}_{I}=0$
does not imply
${\tilde v}^{I}=0$ (nor ${\tilde w}^{I}{\tilde w}_{I}=0$ implies
${\tilde w}^{I}=0$)
because the internal gauge group is $SO(2,1)$ and then there is
the possibility of
having null vectors. In order to show that
$II\subset III$ and $IV\subset III$ we need to
check that any solution to the Gauss law and
${\tilde v}_{I}{\tilde v}^{I}=0$ is a
solution to ${\tilde G}_{I}=0$ and ${\tilde v}_{I}{\tilde w}^{I}=0$.
If ${\tilde v}_{I}=0$ this
is obvious. If ${\tilde v}_{I}$ is null then the internal vectors
$A_{a}^{I}$ are
contained in the null plane orthogonal to ${\tilde v}_{I}$;
the Gauss law, on its
part, tells us that ${\tilde E}^{a}_{I}$ must also be contained in this
null plane;
and thus ${\tilde w}_{I}$ must be proportional to ${\tilde v}_{I}$.
As both of them
are null vectors we conclude that ${\;\;\tilde {\!\!\tilde A}}
\equiv{\tilde v}_{I}{\tilde
w}^{I}=0$. In a similar fashion we can show that $IV\subset III$.
Any point in III
can be shown to be contained in the hypersurfaces defined by some of
these additional
sets of constraints.

We start now studying the possible singularities of I.
To this end we look at the rank
of the matrix defined by the gradients
\begin{equation}
\left[
\begin{array}{c}
d{\tilde v}^{I}\\
d{\tilde G}^{I}
\end{array}
\right]\equiv
\left[
\begin{array}{c}
L\\J
\end{array}
\right]
\left[
\begin{array}{c}
d{\tilde E}^{a}_{I}\\
dA_{aI}
\end{array}
\right]
\label{014bis}
\end{equation}
\noindent where J was defined above and L is the following 3$\times$12 matrix
\begin{equation}
2\left[
\begin{array}{rrrrrrrrrrrr}
0 & 0 & 0 & 0 & 0 & 0 &\!\!0 &
\!\!-A_{2\,3} & \!\!A_{2\,2} & \!\!0 & \!\!A_{1\,3} &
\!\!-A_{1\,2}\\
0 & 0 & 0 & 0 & 0 & 0 &\!\!A_{2\,3} & \!\!0 & \!\!-A_{2\,1} &
\!\!-A_{1\,3} &\!\!0
&A_{1\,1} \\
0 & 0 & 0 & 0 & 0 & 0 & \!\!-A_{2\,2} & \!\!A_{2\,1} & \!\!0 & \!\!A_{1\,2} &
\!\!-A_{1\,1} & \!\!0
\end{array}
\right]
\label{015}
\end{equation}
\noindent by changing coordinates in $\Gamma$ as we did above we see
that as soon as
one of the $A_{aI}$ is non-zero the rank of the matrix in (26) will be
maximal (four). If $A_{aI}=0$ then
the rank can be, at most, three. We see then that the possible
singularities of I
appear when $A_{aI}=0$. Notice that the rank can be maximal for
configurations of the
fields that are singularities of the hypersurface defined by the
Gauss law alone. The
analysis of the singularities in V is completely parallel.
We find that, in this
case, the singular configurations correspond to ${\tilde E}^{a}_{I}=0$.
With this
information we can already see that we have indeed type {\bf d}
singularities in the
full constraint hypersurface III because this points are
intersections of  I and V at
points where these new systems of constraints define smooth
hypersurfaces (and
consequently, they are type {\bf 1}). The same is
true for configurations of type {\bf a} such that not both A's
or both ${\tilde E}$'s are
zero. Let us consider now the possible singularities of II. We have now
\begin{equation}
\left[
\begin{array}{c}
d{\;\;\tilde{\!\!\tilde M}}\\
d{\tilde G}^{I}
\end{array}
\right]\equiv
\left[
\begin{array}{c}
M\\J
\end{array}
\right]
\left[
\begin{array}{c}
d{\tilde E}^{a}_{I}\\
dA_{aI}
\end{array}
\right]
\label{015bis}
\end{equation}
If ${\tilde v}^I=0$ the rank will be at most three and
we have possible singularities.
If ${\tilde v}^I\neq 0$ the rank is easily seen to be four.
This means that the
singularities of II (${\tilde v}^2=0$) are
all contained in I (${\tilde v}^{I}=0$). In
a similar way we show that the singularities in IV
are all contained in V. With this
information we go back to fig. 1. It is straightforward
to see that type {\bf e}
singularities lie at the intersections of ${\tilde v}^2=0$
and ${\tilde w}^{I}=0$ (so
they are type {\bf 1}); with ${\tilde v}^{I}\neq 0$.
We see that these configurations  correspond to
intersections of II and V. Furthermore,
at these points these last two hypersurfaces
are non-singular and thus we conclude
that {\bf e} are genuine singularities of III. A
parallel reasoning applies to type {\bf f}.
The only case that we have not been able to solve
by using these arguments is that of types {\bf b} and {\bf c}
and those configurations
of type {\bf a} with $A_{aI}=0$ or ${\tilde E}^{a}_{I}=0$.
To solve this issue we need to
study the tangent space of III in the vicinity of these points.

Let us prove now that configurations of type {\bf b} are real
singularities of the
constraint hypersurface by showing that the tangent space to the constraint
hypersurface is not defined as in the example of the cone discussed
in section II. Notice that we can take both $A_{1}$ and $A_{2}$ different
from zero and linearly independent because otherwise we would have a
type {\bf a}
singularity. In order to build the required family of curves we write
\begin{eqnarray}
& & {\hat A}_{1\;I}\equiv A_{1\;I}(\epsilon)=A_{1\;I}+\epsilon_{I}\nonumber\\
& & {\hat A}_{2\;I}\equiv A_{2\;I}(\epsilon)=A_{2\;I}+\lambda_{I}\nonumber\\
& & {\hat
E}^{1}_{I}\equiv{\tilde
E}^{1}_{I}(\rho,\epsilon,\sigma,\lambda)=\rho(A_{1\;I}+\epsilon_{I})
+\sigma(A_{2\;I}+\lambda_{I})
\label{1001}\\
& & {\hat E}^{2}_{I}\equiv{\tilde
E}^{2}_{I}(\rho,\epsilon,\sigma,\lambda)=\mu(A_{1\;I}+\epsilon_{I})
+\tau(A_{2\;I}+\lambda_{I})\nonumber
\end{eqnarray}
where $\epsilon_{I}$, $\lambda_{I}$, $\rho$, $\sigma$, $\mu$, and $\tau$ are
parameters such that when they are zero the configuration (\ref{1001})
reduces to the
singularity. In order to satisfy the constraints we must impose some
conditions on the
parameters appearing in (\ref{1001}). The scalar constraint tells
us that, at least
for small arbitrary values of the parameters $\epsilon_{I}$ and $\lambda_{I}$,
${\hat E}^{1}_{I}$ and ${\hat E}^{2}_{I}$ must be
linearly dependent, i.e. $\rho \tau-\mu\sigma=0$.
The Gauss law, on the other hand,
gives the condition
\begin{equation}
(\sigma-\mu)\epsilon_{I}^{\;\;JK}(A_{1\;J}+\epsilon_{J})(A_{2\;K}+
\lambda_{K})=0
\Rightarrow \sigma=\mu
\label{uut}
\end{equation}
so that (\ref{1001}) becomes
\begin{eqnarray}
& & {\hat A}_{1\;I}=A_{1\;I}+\epsilon_{I}\nonumber\\
& & {\hat A}_{2\;I}=A_{2\;I}+\lambda_{I}\nonumber\\
& & {\hat E}^{1}_{I}=\rho(A_{1\;I}+\epsilon_{I})+\sigma(A_{2\;I}+\lambda_{I})
\label{1002}\\
& & {\hat E}^{2}_{I}=\sigma(A_{1\;I}+\epsilon_{I})+\tau(A_{2\;I}+
\lambda_{I})\nonumber
\end{eqnarray}
with the additional condition $\rho \tau-\sigma^{2}=0$
(which is the equation of a
cone). At this point it is not even
necessary to explicitly write down the tangent vectors to the family
of curves
obtained by setting all the
parameters but one equal to zero and differentiating with
respect to the remaining non-zero parameter
because we can see that in the vicinity of a type {\bf b} point the
constraint hypersurface
has the topology of the direct product of a two-dimensional cone and
${\rm I\!R^6}$.
Equation (\ref{1002}) together with $\rho \tau-\sigma^{2}=0$
is the general solution
to the constraints in the vicinity of a type {\bf b}
singularity only if the plane that
contains  $A_{aI}$ is not null; if it is null then the
argument presented above still
proves
that we have a singularity but the previous solution is not the most
general one. A
completely parallel argument applies to type {\bf c} configurations.
As they cannot be
described as intersections of smooth manifolds they are type {\bf 2}.
In order to prove
that  type {\bf a} singularities with $A_{aI}=0$, ${\tilde E}^{a}_{I}=0$
or both are
singularities we use the same kind of ideas. In the
case in which both $A_{aI}=0$
and ${\tilde E}^{a}_{I}=0$ we choose the set of curves
\begin{equation}
\begin{array}{ll}
\left\{
\begin{array}{l}
A_{1\;I}=\lambda_{I}\\
A_{2\;I}=\mu_{I}\\
{\tilde E}^{1}_{I}=0\\
{\tilde E}^{2}_{I}=0
\end{array}
\right.&
\hspace{2cm}
\left\{
\begin{array}{l}
A_{1\;I}=0\\
A_{2\;I}=0\\
{\tilde E}^{1}_{I}=\rho_{I}\\
{\tilde E}^{2}_{I}=\sigma_{I}
\end{array}
\right.
\end{array}
\label{1003}
\end{equation}
where $\lambda_{I}$, $\mu_{I}$, $\rho_{I}$, $\sigma_{I}$ are
parameters. Obviously we get a 12 dimensional vector space from
the tangent vectors obtained by putting all the parameters but
one to zero and
differentiating with respect to the parameter left.
If ${\tilde E}^{a}_{I}=0$ but
$A_{aI}\neq 0$ we choose ($A_{aI}\equiv a_{a}\tau_{I}$ with $a_{2}\neq 0$)
\begin{equation}
\begin{array}{ll}
\left\{
\begin{array}{l}
A_{1\;I}=a_{1}\tau_{I}+\lambda_{I}\\
A_{2\;I}=a_{2}\tau_{I}+\epsilon_{I}\\
{\tilde E}^{1}_{I}=0\\
{\tilde E}^{2}_{I}=0
\end{array}
\right.&
\hspace{2cm}
\left\{
\begin{array}{l}
A_{1\;I}=a_{1} \tau_{I}\\
A_{2\;I}=a_{2} \tau_{I}\\
{\tilde E}^{1}_{I}=\rho_{I}\\
{\tilde E}^{2}_{I}=\frac{1}{a_{2}}(\xi\tau_{I}-a_{1}\rho_{I})
\end{array}
\right.
\end{array}
\label{1004}
\end{equation}
where $\tau_{I}$ is a fixed internal vector in the direction of $A_{aI}$ and
$\rho_{I}$, $\lambda_{I}$, $\epsilon_{I}$, and $\xi$ are parameters. Notice
that now we do not have the kind of conical
singularity that we found before because we do not have two linearly
independent
internal vectors. The tangent
vectors to the previous set of curves span a 10-dimensional vector space
thus proving
that these configurations are also singular. Although these
singularities lie at
intersections of some of the other sectors of the theory
we classify them as type {\bf
2} because at these points the surfaces that describe these other sectors are
themselves singular. As we
will show in the next section there is a natural way of
defining dynamics for configurations that lie at type {\bf 1} singular points.

A diagram representing the mutual relationships between the
different sectors of the
constraint hypersurface is shown in fig. 3
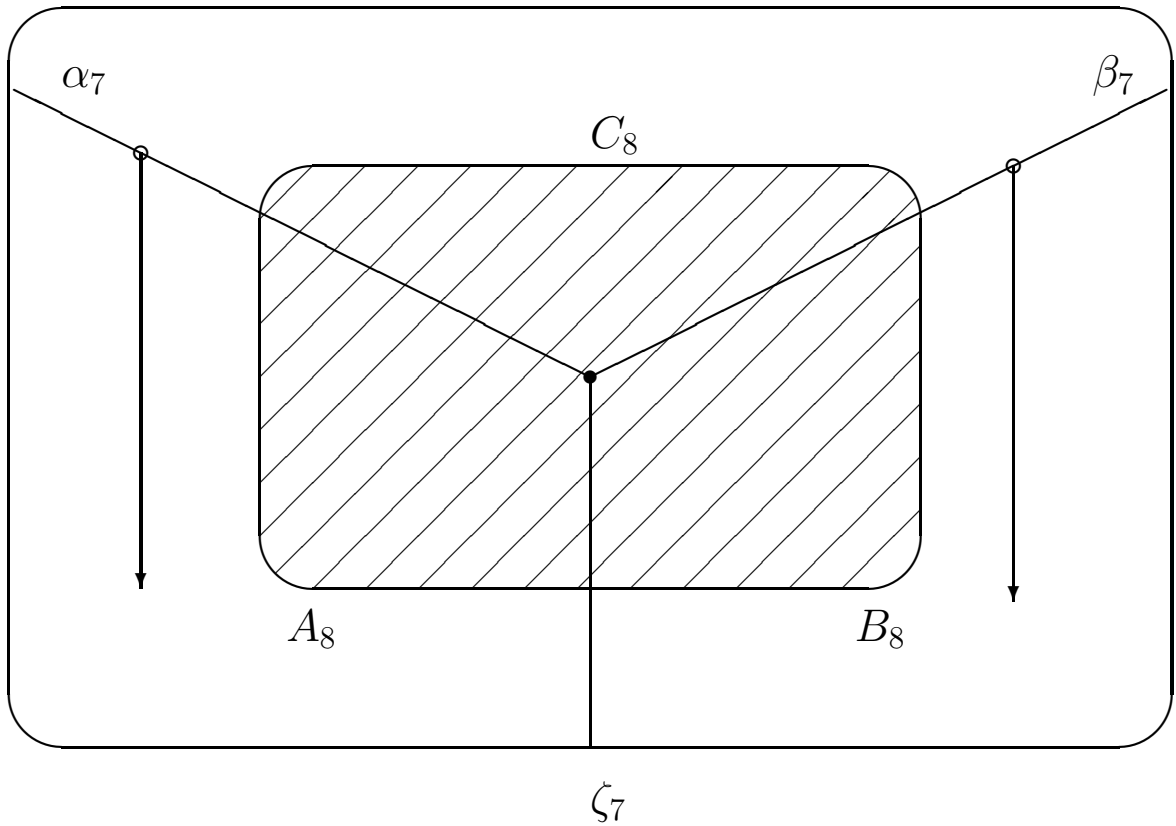
\begin{figure}
\begin{picture}(500,340)(30,20)
\thicklines
\Large
\put(250,200){\oval(440,280)}
\put(250,200){\oval(250,160)}
\put(250,200){\line(-2,1){218}}
\put(250,200){\line(2,1){218}}
\put(250,200){\line(0,-1){140}}
\put(250,200){\circle*{5}}
\put(350,100){$B_{8}$}
\put(135,100){$A_{8}$}
\put(250,287){$C_{8}$}
\put(250,35){$\zeta_{7}$}
\put(440,310){$\beta_{7}$}
\put(50,310){$\alpha_{7}$}
\put(410,280){\circle{5}}
\put(80,285){\circle{5}}
\put(410,280){\vector(0,-1){165}}
\put(80,285){\vector(0,-1){165}}
\thinlines
\put(125,200){\line(1,1){80}}
\put(125,220){\line(1,1){60}}
\put(125,240){\line(1,1){40}}
\put(125,260){\line(1,1){20}}
\put(125,180){\line(1,1){100}}
\put(125,160){\line(1,1){120}}
\put(125,140){\line(1,1){140}}
\put(131,126){\line(1,1){154}}
\put(145,120){\line(1,1){160}}
\put(165,120){\line(1,1){160}}
\put(185,120){\line(1,1){160}}
\put(205,120){\line(1,1){158}}
\put(225,120){\line(1,1){148}}
\put(245,120){\line(1,1){130}}
\put(265,120){\line(1,1){110}}
\put(285,120){\line(1,1){90}}
\put(305,120){\line(1,1){70}}
\put(325,120){\line(1,1){50}}
\put(345,120){\line(1,1){30}}
\end{picture}

\caption{The constraint hypersurface. The subindices in the
labels of each region
denote
their dimensionality. The two arrows represent type {\bf b} or {\bf c}
singularities and the
overshadowed region represents those points in the constraint
hypersurface accessible
from physical initial data (defined in section IV).}
\end{figure}
The points in each of the regions represented in the figure satisfy
the following
conditions
\begin{equation}
\begin{array}{llllll}
 B_{8} & {\tilde v}_{I}=0 & {\tilde G}_{I}=0 & {\tilde w}_{I}\neq 0 &
{\rm not\;\;null}\\
A_{8} & {\tilde w}_{I}=0 & {\tilde G}_{I}=0 & {\tilde v}_{I}\neq 0 &
{\rm not\;\;null}\\
C_{8} & {\tilde w}^{2}=0 & {\tilde v}^{2}=0 & {\tilde v}_{I}\neq 0 &
{\tilde w}_{I}\neq 0\;\;\;{\tilde G}_{I}=0\\
\zeta_{7} & {\tilde w}_{I}=0 & {\tilde v}_{I}=0 & {\tilde G}_{I}=0 & &\\
\beta_{7} &{\tilde w}^{2}=0 & {\tilde w}_{I}\neq 0 & {\tilde v}_{I}=0 &
{\tilde G}_{I}=0 &\\
\alpha_{7} & {\tilde v}^{2}=0 & {\tilde v}_{I}\neq 0 & {\tilde w}_{I}=0 &
{\tilde G}_{I}=0 &
\end{array}
\nonumber
\end{equation}
Type {\bf a} singularities are contained in $\zeta_{7}$, type {\bf b}
singularities are
contained in $A_{8}$ and $\alpha_{7}$, type {\bf c} singularities
are contained in $B_{8}$ and
$\beta_{7}$, type {\bf d} singularities are contained in $\zeta_{7}$, type
{\bf e}
singularities are contained in $\alpha_{7}$ and type {\bf f}
singularities are contained in
$\beta_{7}$. The constraint ${\tilde w}^{I}=0$, together with the Gauss law,
describes
points in $A_{8}$, $\zeta_{7}$, and $\alpha_{7}$,  ${\tilde v}^{I}=0$
points in
$B_{8}$, $\zeta_{7}$, and $\beta_{7}$, ${\tilde w}^{2}=0$
points in $C_{8}$ and
$\beta_{7}$, and finally ${\tilde v}^{2}=0$ points in $C_{8}$ and
$\alpha_{7}$.

\section{Dynamics}

In this section
we will concentrate on the study of the evolution of ``physical" initial
data. We will call ``physical''
 those initial data that satisfy the following
two conditions: (i) $({\tilde E}^{a}_{I}{\tilde E}^{bI})$ is
nondegenerate and of
 ++ signature and (ii) the data are non-singular points
of the constraint hypersurface.
We start by proving that for physical initial data the quantity
${\tilde E}^{a}_{I} A_{a}^{I}
\equiv E\cdot A$ is non-zero
(The importance of this fact is that, as we will show  below,
this quantity is conserved under the evolution defined by all the
previous sets of
constraints. This is very useful when discussing dynamics.). The fact
that the (densitized) 2-metric
${\tilde E}^{a}_{I}{\tilde E}^{bI}$ is non-degenerate implies that
${\tilde  E}^{1}_{I}$ and
${\tilde  E}^{2}_{I}$ are not collinear and not contained in a null plane.
This implies that,
necessarily $A_{1\;I}$ and $A_{2\;I}$ are collinear and contained in the
plane spanned by
${\tilde  E}^{a}_{I}$. Let us write
\begin{equation}
\begin{array}{l}
{\tilde E}^{1}_{I}=e_{1} \tau_{I}+\varrho_{1}\mu_{I}\\
{\tilde E}^{2}_{I}=e_{2} \tau_{I}+\varrho_{2}\mu_{I}\\
A_{1\;I}=a_{1}\tau_{I}\\
A_{2\;I}=a_{2}\tau_{I}
\end{array}
\label{1005}
\end{equation}
where $\tau_{I}$ and $\mu_{I}$ are orthonormal vectors. The Gauss law implies
$a_{1}\varrho_{1}+a_{2}\varrho_{2}=0$ and the scalar constraint
is immediately satisfied. The
non-degeneracy condition of the metric is $e_{1}\varrho_{2}-e_{2}\varrho_{1}
\neq 0$. If we
suppose that $A\cdot E=0$ ($a_{1}e_{1}+a_{2}e_{2}=0$) we must have $a_{1}=0$
and $a_{2}=0$
($e_{1}\varrho_{2}-e_{2}\varrho_{1}\neq 0$ implies that this is the only
solution to
$a_{1}e_{1}+a_{2}e_{2}=0$ and  $a_{1}\varrho_{1}+a_{2}\varrho_{2}=0$).
 As we have seen before,
points for which $A_{aI}=0$ are singularities of the constraint
hypersurface and hence
they
are not physical data; so we conclude that physical configurations
must always satisfy
$A\cdot E\neq 0$. Let us prove now that $A\cdot E$ is conserved.
As it is gauge
invariant we have to consider only its evolution
under $\;\;{\tilde {\!\!\tilde A}}=0$,
$\;\;{\tilde {\!\!\tilde F}}=0$, $\;\;{\tilde {\!\!\tilde M}}=0$,
${\tilde v}_{I}=0$, and
${\tilde w}_{I}=0$. By using  the following Poisson brackets
\begin{equation}
\begin{array}{ll}
\{{\tilde E}^{a}_{I} A_{a}^{I},A_{b}^{J}\}=-A_{b}^{J}&
\{{\tilde E}^{a}_{I} A_{a}^{I},
{\tilde E}^{b}_{J}\}={\tilde E}^{b}_{J}\\
\{{\tilde E}^{a}_{I} A_{a}^{I}, {\tilde v}_{J}\}=-2 {\tilde v}_{J} &
\{{\tilde E}^{a}_{I} A_{a}^{I}, {\tilde w}_{J}\}=2 {\tilde w}_{J} \\
\{{\tilde E}^{a}_{I} A_{a}^{I}, {\tilde v}^{2}\}=-4{\tilde v}^{2} &
\{{\tilde E}^{a}_{I} A_{a}^{I}, {\tilde w}^{2}\}=4{\tilde w}^{2}\\
\{{\tilde E}^{a}_{I} A_{a}^{I}, {\tilde w}_{I}{\tilde v}^{I}\}=0 &
\end{array}
\label{1006}
\end{equation}
it is easy to show that $A\cdot E$ is a constant of motion for all the above
systems of constraints.
We describe now those singular configurations that can be reached from
physical
initial data. It is straightforward to show that configurations
of types {\bf b}, {\bf
c}, {\bf d} all
have $E\cdot A=0$. This is obvious for {\bf b} and {\bf c}.
For {\bf d} we write ${\tilde
E}^{1}_{I}=e_{1}\tau_{I}$, ${\tilde E}^{2}_{I}=e_{2}\tau_{I}$,
$A_{1}=a_{1}\mu_{I}$,
and $A_{2}=a_{2}\mu_{I}$ with $\tau_{I}$ and $\mu_{I}$ linearly
independent. The Gauss
law implies $e_{1}a_{1}+e_{2}a_{2}=0$ and thus $E\cdot A=0$.
We conclude that the only
singular configurations that we can reach from
physical initial data are type {\bf
a}, {\bf e} or {\bf f}. All of them are type {\bf 1}.
The overshadowed region in fig. 3 represents the part
of the constraint hypersurface
accessible from physical
initial data. Notice that type {\bf b} and {\bf c} singularities are
excluded from this region. The accessible singularities in
$\beta_{7}$ are all type {\bf f}, those in $\alpha_{7}$ are type {\bf e},
and those in
$\zeta_{7}$ are type {\bf a} (with $E\cdot A\neq0$).

Thus we have shown that
for physical initial data,
all the singularities accessible through evolution
are at intersections of smooth hypersurfaces in the
phase space. We define evolution  through these
singularities
simply by using the alternative evolutions defined by the functions
that describe the smooth manifolds that intersect to create the singularities.
 We discuss
this in detail now.
The evolution equations generated by the constraint functions (suppressing
the evolution generated by the Gauss law constraint ${\tilde G}_{I}=0$) are
\begin{equation}
\begin{array}{ll}
    \left\{\begin{array}{l}
        {\tilde v}^{I}=0;\;\;{\tilde G}_{I}=0\\
        {\dot{\tilde E}^{a}}_{I}=2 \epsilon_{IJK}
{\tilde \eta}^{ab}A_{b}^{J} \lambda^{K}\\
        {\dot A_{aI}}=0\\
        {\dot{\tilde w}}^{I}=4\left[ ({\tilde E}^{a}_{J}\lambda^{J})
A_{a}^{I}-(E\cdot
           A)\lambda^{I}\right]
    \end{array}\right. &
    \left\{\begin{array}{l}
        {\tilde w}^{I}=0;\;{\tilde G}_{I}=0\\
        {\dot{\tilde E}^{a}}_{I}=0 \\
        {\dot A_{aI}}=2 \epsilon^{IJK} \alpha_{J}\eut_{ab}{\tilde E}^{b}_{K}\\
        {\dot{\tilde v}}^{I}=4\left[(E\cdot A)\alpha^{I}-{\tilde E}^{aI}
(A_{a}^{J}
          \alpha_{J})\right]
    \end{array}\right.\\
    \left\{\begin{array}{l}
        {\tilde v}^{2}=0;\;\;{\tilde G}_{I}=0\\
        {\dot{\tilde E}^{a}}_{I}=-4 \nut\epsilon_{IJK}{\tilde v}^{J}
A_{b}^{K}\\
        {\dot A_{aI}}=0\\
        {\dot{\tilde w}}^{I}=-8\nut\left[{\tilde v}^{I}(E\cdot A)\right]
    \end{array}\right. &
    \left\{\begin{array}{l}
        {\tilde w}^{2}=0;\;\;{\tilde G}_{I}=0\\
        {\dot{\tilde E}^{a}}_{I}=0\\
        {\dot A}_{aI}=4\mut\eut_{ab}\epsilon^{IJK}{\tilde w}_{J}
{\tilde E}^{b}_{K}\\
        {\dot{\tilde v}}^{I}=8\mut{\tilde w}^{I}(E\cdot A)
    \end{array}\right.
\end{array}
\label{r1}
\end{equation}
where the dot represents the derivative with respect to some parameter `t'.
By evolution
we mean motion generated by a constraint function (obtained from a
constraint by
multiplying it by a suitable Lagrange multiplier) via Poisson brackets.
In (\ref{r1})
$\alpha_{I}$, $\lambda_{I}$, $\mut$, and $\nut$ are (t-dependent)
Lagrange multipliers. The
equations above treat $A_{aI}$ and ${\tilde E}^{a}_{I}$ symmetrically
so we can learn about
some of the sectors by studying the others. In $C_{8}$
we are allowed to use either
${\tilde v}^{2}=0$ or ${\tilde w}^{2}=0$ (together with ${\tilde G}_{I}=0$).
It is
straightforward to check that, as long as ${\tilde v}^{I}\neq 0$ and ${\tilde
w}^{I}\neq 0$ both sets of evolutions are equivalent, as expected
from the fact that
both functions define the same part of the constraint hypersurface.

The result we set out to prove in the remainder of this section is that,
in a precise sense, for all points gauge equivalent to physical data, the
reduced phase space of the Ashtekar and Witten formulations coincide.
Notice that
the possibility of different reduced phase spaces for
(gauge equivalence classes of) physical
data in the Ashtekar and the Witten formulations arises because of the
various intersections present in the constraint surface of the Ashtekar
theory. Thus, our aim is to show that these intersections do not alter the
reduced phase space.

Before giving the proof in full detail, we first state the main points
below:

\noindent (1) We first show that every physical data point is gauge
equivalent
to some point in $\zeta_7$.\\
\noindent (2) Next, we show  that the intersection of the
gauge orbits of $A_8$
with the physically relevant part of $\zeta_7$ does not lead to
identifications of points in $\zeta_7$ which were not already identified
by the gauge orbits of $B_8$\\
\noindent (3) We show that every point in $\zeta_7$ obtained from
physical data is gauge equivalent to certain points in $\alpha_7$ and
$\beta_7$\\
\noindent (4) We show that  (3) implies the gauge identification
of points  within $C_8$ which were hitherto
not identified by gauge transformations
only generated by the constraints defining $C_8$.\\
\noindent (5) However (we also
show that) gauge transformations generated by the constraints defining $C_8$
{\em do not} provide extra identifications of points in $\beta_7$ and
$\alpha_7$ over and above those identifications already made by
gauge transformations generated by the constraints defining $A_8$ and $B_8$.\\

The above shows that for physical data points in
$B_8$, there are no extra (gauge) identifications with other points
within $B_8$ due to the presence of the sectors $A_8$ and $C_8$.
As $B_8$ is exactly the Witten constraint surface we have then
proved the previous statement about the equivalence of the Ashtekar and
the Witten formulations.

Physically relevant data (which lie in $B_{8}$) have ${\tilde w}^{I}$
time-like. We can
show that by evolving with ${\tilde v}^{I}=0$ we can make ${\tilde w}^{I}=0$
(that is, reach the singular region $\zeta_{7}$). Indeed, choosing
$\lambda^{I}(t)={\tilde {w}}^{I}(0)$ and taking into account that
under the evolution
given by ${\tilde v}^{I}=0$ the quantity ${\tilde E}^{a}_{I} A_{a}^{J}$
is conserved,
we have
\begin{equation}
\dot{\tilde w}_{I}=-4(E\cdot A){\tilde w}_{I}(0)\Rightarrow {\tilde w}_{I}(t)=
{\tilde w}_{I}(0)\left[1-4(E\cdot A)t\right]
\label{erere}
\end{equation}
so if $t=\frac{1}{4(E\cdot A)}$ we hit the singularity at $\zeta_{7}$.
As we can see,
it is possible to connect non-degenerate
metrics to degenerate ones for initial data
such that $E\cdot A\neq0$; this proves point 1.

We have already seen in the previous section  that the
constraint hypersurface is
singular in several regions. The sectors for which $E\cdot A\neq0$
 have the nice property of being individually non-singular,
the singularities of the
full constraint hypersurface
are just intersections between the different non-singular
sectors. Let us comment on such possibilities by looking at the following
 example. Suppose that we
take the union of I and V
as our constraint hypersurface and impose $E\cdot A\neq0$. In spite of the
presence of a type {\bf 1}
singularity the reduced phase space may still be well defined and not
inherit any non-smooth properties of the intersection region if those motions
generated
by  ${\tilde v}^{I}=0$ and ${\tilde G}^{I}=0$ which connect points
with ${\tilde
w}^{I}=0$ do not provide extra identifications in the ${\tilde w}^{I}=0$
sector over
and above those provided by motions generated by ${\tilde w}^{I}=0$
and ${\tilde G}^{I}=0$
themselves (and vice versa for motions generated by ${\tilde w}^{I}=0$ and
${\tilde G}^{I}=0$ which connect points with ${\tilde v}^{I}=0$).
In such a situation
the reduced phase space is exactly the same as that corresponding only to
${\tilde v}^{I}=0$ and ${\tilde G}^{I}=0$. For those singularities that are
intersections of smooth hypersurfaces we can define a finite number of
alternative evolutions by restricting ourselves to each smooth hypersurface
 separately.
It is possible to have more complicated behaviors than in the above example
(such as in the
intersections of I, IV and II, V, as shown below) and still obtain a well
behaved reduced phase space.
For singularities that are not intersections of this type (for example
conical singularities) the issue of how to define evolution may be much
more involved
and it is not clear if the evolution can be defined in those cases.

We proved above that by evolving physical initial data we can always reach
$\zeta_{7}$.
Let us  consider now initial data on $\zeta_{7}$ and discuss point 2.
Let us write
\begin{equation}
\begin{array}{l}
A_{aI}=a_{a}x_{I}\\
{\tilde E}^{a}_{I}=e^{a} x_{I}
\end{array}
\label{r2}
\end{equation}
with $x_{I}$ an arbitrary unit space-like vector
and $E\cdot A =e^{a}a_{a}\neq 0$.
Notice that within $B_{8}$, $A_{aI}$ must have the form
shown in (\ref{r2}). The
fact that a physical ${\tilde E}^{a}_{I}$ in $B_{8}$ has to give a
(+,+) signature metric tells us that the
plane
containing $A_{aI}$ and ${\tilde E}^{a}_{I}$ must be spatial; this implies
that $x_{I}$ is  space-like. Let us consider first the
evolution given by ${\tilde
v}^{I}=0$.
Solving the evolution equations we get
\begin{equation}
\begin{array}{l}
A_{aI}(t)=A_{aI}(0)=a_{a}x_{I}\\
{\tilde E}^{a}_{I}=e^{a}x_{I}+ 2 \epsilon_{IJK}{\tilde
\eta}^{ab}a_{b}x^{J}\beta^{K}(t)\\
{\tilde w}_{I}=4(e^{a}a_{a})[-\delta_{I}^{J}+x_{I}x^{J}]\beta_{J}(t)\\
\beta_{I}(t)\equiv\int_{0}^{t}\lambda_{I}(\tau)d\tau
\end{array}
\label{r3}
\end{equation}
If we want to stay in $\zeta_{7}$ with this evolution we must demand ${\tilde
w}_{I}(t)=0$,
which implies $\beta_{I}(t)=\beta(t) x_{I}$. Substituting this into the
equation for ${\tilde E}^{a}_{I}(t)$ we see that ${\tilde E}^{a}_{I}(t)=
{\tilde E}^{a}_{I}(0)=e^{a}x_{I}$ i.e. it is impossible
to evolve within $\zeta_{7}$ by
using ${\tilde v}^{I}=0$.

Suppose now that we want to know if it is possible to hit
$\beta_{7}$ by evolving these initial data (point 3). To this
end we must require ${\tilde
w}^{2}=0$, ${\tilde w}_{I}\neq 0$. We have then
\begin{equation}
{\tilde w}^{2}=16(e^{a}a_{a})^2(\delta_{I}^{J}-x_{I}x^{J})
\beta^{I}(t)\beta_{J}(t)=0
\label{r4}
\end{equation}
The general solution to the previous equation is of the form
$\beta_{I}=\alpha x_{I}+\gamma l^{\pm}_{I}$ where the
two null vectors $l^{\pm}_{I}$
are defined by  $l^{\pm}_{I}=t_{I}\pm y_{I}$, and $(t_{I},x_{I},y_{I})$ is an
orthonormal basis such that $t_{I}$ is time-like. For this $\beta_{I}$ we have
\begin{equation}
\begin{array}{l}
{\tilde E}^{a}_{I}=e^{a}x_{I}\mp 2{\tilde \eta}^{ab}a_{b}\gamma l^{\pm}_{I}\\
A_{aI}=a_{a}x_{I}\\
{\tilde w}_{I}=-4\gamma(e^{a}a_{a})l^{\pm}_{I}
\end{array}
\label{r5}
\end{equation}
so we can indeed reach the singularity at $\beta_{7}$.
The previous result shows
an interesting property of the
dynamics (point 4): there are field configurations on $\beta_{7}$
that are not connected by $SO(2,1)$ transformations
(nor the evolution defined in
$C_{8}$)  but are gauge equivalent under the
evolution generated by ${\tilde v}^{I}=0$.

 In the previous computation we have not
included the motions generated by the Gauss law. If
we do so we obtain the following
set of equations
\begin{equation}
\begin{array}{l}
\dot{\tilde E}^{a}_{I}=2\epsilon_{IJK}{\tilde
\eta}^{ab}A_{b}^{J}\lambda^{K}+\epsilon_{IJK}\theta^{J}{\tilde E}^{aK}\\
\dot A_{aI}=\epsilon_{IJK}\theta^{J}A_{a}^{K}\\
\dot{\tilde w}_{I}=4\left[({\tilde E}^{A}_{J}\lambda^{J})A_{a}^{I}-(E\cdot
A)\lambda^{I}\right]+\epsilon_{IJK}\theta^{J}{\tilde w}^{K}
\end{array}
\label{r6}
\end{equation}
where $\theta_{I}$
is an additional Lagrange multiplier.
We can always integrate the equation for $A_{aI}$ to get
\begin{equation}
A_{aI}(t)=\Lambda^{J}_{\;\;I}A_{aJ}(0)
\label{r7}
\end{equation}
where $\Lambda_{\;\;I}^{J}$  is a finite $SO(2,1)$ transformation such that
\begin{equation}
\dot \Lambda_{\;\;I}^{J}=
\epsilon^{KL}_{\;\;\;\;\;I}\theta_{K}(t)\Lambda^{J}_{\;\;L}(t)
\label{r8}
\end{equation}
Defining ${\hat E}^{aI}=\Lambda^{I}_{\;\;J}{\tilde E}^{aJ}$,
${\hat \lambda}^{I}=
\Lambda^{I}_{\;\;J}\lambda^{J}$,
${\hat w}^{I}=\Lambda^{I}_{\;\;J}{\tilde w}^{J}$ and
using the facts that $(\Lambda^{I}_{\;\;J})^{-1}=\Lambda_{J}^{\;\;I}$
and $\epsilon^{IJK}=\epsilon^{LMN}
\Lambda^{I}_{\;\;L}\Lambda^{J}_{\;\;M}\Lambda^{K}_{\;\;N}$ we get the
following equations for ${\tilde E}^{a}_{I}$ and ${\tilde w}_{I}$.
\begin{equation}
\begin{array}{l}
\dot{{\hat E}^{a}_{I}}=2 \epsilon_{IJK}{\tilde
\eta}^{ab}A_{b}^{J}(0){\hat \lambda}^{K}\\
\dot {\hat w}_{I}=4\left[({\hat E}^{a}_{J}{\hat \lambda}^{J})A_{aI}(0)-
(E\cdot A){\hat
\lambda}^{J}\right]
\end{array}
\label{r9}
\end{equation}
These equations have the same form as before, so the same analysis
gives now the
following result.
If we impose ${\hat w}_{I}(t)=0$ we get ${\hat E}^{a}_{I}(t)={\tilde
E}^{a}_{I}(0)=e^{a}x_{I}\Rightarrow
\Lambda_{I}^{\;\;J}(t){\tilde E}^{a}_{J}(t)={\tilde
E}^{a}_{I}(0)$ so that,
the resulting motion is equivalent to an $SO(2,1)$ gauge
transformation. The same analysis can
be done with the requirement ${\hat w}^{2}=0$ to get
configurations that are $SO(2,1)$ gauge equivalent to (\ref{r5}).

Let us consider now the evolution
defined by ${\tilde w}^{I}=0$ given by (\ref{r1}).
Solving the evolution equations we get
\begin{equation}
\begin{array}{l}
{\tilde E}^{a}_{I}(t)={\tilde E}^{a}_{I}(0)=e^{a}x_{I}\\
A_{a}^{I}(t)=a_{a}x_{I}-2\epsilon^{IJK}\eut_{ab}e^{b}x_{J}\zeta_{K}(t)\\
{\tilde v}_{I}(t)=4(e^{a}a_{a})\left[\delta_{I}^{J}-x_{I}x^{J}\right]
\zeta_{J}\\
\zeta_{I}(t)\equiv\int_{0}^{t}\zeta{I}(\tau)d\tau
\end{array}
\label{r10}
\end{equation}
As before it is
impossible to evolve within $\zeta_{7}$ by using ${\tilde w}^{I}=0$. Also,
 we can take into account the $SO(2,1)$
gauge transformations as we did before.

When we
hit the singularity $\alpha_{7}$ we do it at points of the form
\begin{equation}
\begin{array}{l}
{\tilde E}^{a}_{I}=e^{a}x_{I}\\
A_{aI}=a_{a}x_{I}\pm 2\eut_{ab}e^{b}\rho l^{\pm}_{I}\\
{\tilde v}_{I}=4\rho(e^{a}a_{a})l^{\pm}_{I}
\end{array}
\label{r11}
\end{equation}
The argument is
essentially the
same if we allow for $SO(2,1)$ evolution too. The last remaining step
(point 5)
to prove the consistency of the
evolution is to show that the configurations that we
find at $\beta_{7}$ and $\alpha_{7}$
are gauge related under the evolution generated by
the constraints in $C_{8}$.
To this end we evolve (\ref{r5}) with ${\tilde w}^{2}=0$
and (\ref{r11}) with ${\tilde v}^{2}=0$ to get
\begin{equation}
\begin{array}{l}
A_{aI}=a_{a}x_{I}\mp 16 B\gamma(e^{c}a_{c})\eut_{ab}e^{b}l^{\pm}_{I}\\
{\tilde E}^{a}_{I}=e^{a}x_{I}\mp 2{\tilde\eta}^{ab}a_{b}\gamma l^{\pm}_{I}\\
\end{array}
\label{r16}
\end{equation}
\begin{equation}
\begin{array}{l}
A_{aI}=a_{a}x_{I}\pm 2\eut_{ab}e^{b}\rho l^{\pm}_{I}\\
{\tilde E}^{a}_{I}
=e^{a}x_{I}\mp16\rho A(e^{c}a_{c}){\tilde \eta}^{ab}a_{b}l^{\pm}_{I}
\end{array}
\label{r17}
\end{equation}
where $A=\int_{0}^{t}\nut(\tau)d\tau$,
$B=\int_{0}^{t}\mut(\tau)d\tau$. As we can see, it
is always possible to choose
the Lagrange multipliers in such a way that the $A_{aI}$
obtained by evolving from $\beta_{7}$
and $\alpha_{7}$ coincide and also the ${\tilde
E}^{a}_{I}$.
This is true for upper and lower signs in $l^{\pm}_{I}$ separately.
However, it is
not possible
to connect configurations with different null vectors $l^{\pm}_{I}$ by
evolving through this region.
In the previous argument the evolutions in $A_{8}$ and
$B_{8}$ were
required to reach the singularities at $\alpha_{7}$ and $\beta_{7}$ but
were, otherwise,
arbitrary. It was then found that it is possible to find Lagrange
multipliers such
that configurations in both singular regions were appropriately
connected.

Finally,
we have also
examined  the following evolution of points in $\zeta_7$ which are
gauge related
to physical initial data:
We allow arbitrary evolution of such points through
$A_{8}$ subject to the condition that we hit $\alpha_7$. From $\alpha_7$ we
allow arbitrary evolution in $C_8$
subject to  the condition that we hit $\beta_7$. We have been able to show,
by integrating out the equations of motion, that we can ``close the
orbits" in the remaining region $B_8$ i.e. the point we obtain on $\beta_7$
is gauge related by motions through  $B_8$
to the point in $\zeta_7$ we started out with.
The same result is true for interchange of $A_8$ with $B_8$ and $\alpha_7$
with $\beta_7$.

All  the previous arguments go through also if we
take into
account the evolution generated by the Gauss law. We conclude then that,
even in the
presence of extra sectors, this homogeneous model has the same reduced
phase space as the Witten formulation.
Maybe a similar statement can be made in the non-homogeneous case as
well.
We have not studied the non-physical initial data. In this case it may be
possible
that the reduced phase spaces of the Ashtekar and Witten formulations are
different.

\section{Conclusions}

Let us first summarize our results.
We have studied a homogeneous reduction of 2+1 dimensional gravity in the
Ashtekar
formulation using a `geometric viewpoint'.
The constraint hypersurface is  a complicated 8-dimensional
object embedded in the 12-dimensional phase space. It is possible to
show that there are several singular regions in it.
By restricting ourselves to the
evolution of physical initial data we have shown
that the singularities
that can be reached from such data are of a ``mild" type --they are
 intersections
of pairs of smooth 8-dimensional manifolds--.
This allows  a definition of dynamics
through such singularities. When the gauge orbits hit
these singular configurations there are only two
possible alternative ways to continue the evolution obtained
by using the two sets of
constraints
defining the two intersecting manifolds.
The key issue at this point is to check that
there are not extra gauge identifications
produced by the global structure of the
constraint hypersurface. As shown in the previous section no problems arise
(Note that the analogs of points where the connection identically vanishes
were a source of pathology in the study of the reduced phase space of the
Witten formulation (without a homogeneity ansatz) in \cite{jorma}).

A  similar analysis to the one presented in this work
for the non-homogeneous case would be
very interesting but we expect the
technical details to be more involved than the simple arguments
presented here.
In particular, it would be nice if some statement of equivalence
(or nonequivalence) of
the physical sectors of the Ashtekar and Witten theories could be made. For
example, is
the infinite dimensional sector of the Ashtekar theory \cite{Madh}
in a pathological part of the constraint surface?  In fact
one may ask as to whether, using our geometrical viewpoint, there is a well
defined physical sector of the theory at all. Also, it would be useful to
see whether the geometric viewpoint gives rise to the existence of extra
gauge orbits in the infinite dimensional sector which reduce the dimension
to a finite number. This is of interest especially because there are
indications \cite{bernie} that it may be that, with certain choices of
admissible wave functions,  the quantum theories of the Ashtekar and Witten
formulations are identical.

It would  be interesting to see whether the geometric viewpoint
indicates that the negative energy sector in the non-compact case
\cite{madthesis} is in a pathological sector of the constraint surface.
In fact due to the similar structure of the constraints in 2+1 and 3+1
dimensions, if this can be done, it may
even have a bearing on the negative energy solutions of
\cite{mad-ve} in the 3+1 theory.

Apart from all this, we have shown that the viewpoint in this paper
has allowed us to deal with dynamical issues related to degenerate metrics,
at least in a cosmological scenario. It would be interesting to see
if we could identify singularities
in 3+1 Bianchi models with degeneracies of the Ashtekar triads and evolve
through  these degeneracies using the techniques in this paper.

{\bf Acknowledgements} The authors want to thank Jorma Louko for an
interesting discussion.
Madhavan Varadarajan is supported by NSF grant PHY-9207225
and J.F.B.G. by the NSF under grant PHY-9396246 and funds
provided by the Pennsylvania
State University.

\end{document}